\documentclass{svproc}

\usepackage[table]{xcolor}
\definecolor{qnowcolor}{HTML}{30A08E}
\usepackage{colortbl}
\usepackage{booktabs}
\usepackage{hhline}
\usepackage{array, makecell}

\usepackage{amsmath,amssymb,amsfonts}
\usepackage{mathrsfs}
\usepackage[scr=dutchcal,calscaled=1]{mathalfa}
\usepackage{braket}

\usepackage{pifont}
\usepackage{hyperref}
\hypersetup{
  colorlinks=true,
  linkcolor=qnowcolor,
  filecolor=qnowcolor,
  urlcolor=qnowcolor,
  citecolor=qnowcolor
}

\usepackage{graphicx}
\usepackage{textcomp}
\usepackage{url}

\usepackage{comment}

\usepackage{orcidlink}

\usepackage[natbibapa]{apacite}
\begin{document}
\mainmatter

\begin{comment}
% Spanish title here
\title{Quantum \texttt{QSAR} para el descubrimiento de fármacos}
\titlerunning{Quantum QSAR}
\end{comment}
% English title herre

\title{Quantum \texttt{QSAR} for drug discovery}

\author{
Alejandro Giraldo\,\thanks{corresponding author:  \email{alejandro@qnow.tech}} \inst{1}\orcidlink{0009-0008-9826-0703} \and
Daniel Ruiz\,\orcidlink{0009-0007-6976-1755} \and \\
Mariano Caruso\, \inst{2,3,4}\orcidlink{0000-0002-7455-1193} \and
Guido Bellomo\,\inst{5}\orcidlink{0000-0001-8213-8270}
}

\authorrunning{A. Giraldo, D. Ruiz, M. Caruso, G. Bellomo} 

\institute{
\href{https://qnow.tech/}{\texttt{QNOW Technologies}}, Delaware, USA\\
\email{alejandro@qnow.tech}, \email{daniel@qnow.tech} \and
\href{https://www.ugr.es/}{\texttt{UGR}}, Granada, Spain \and
\href{https://www.unir.net/}{\texttt{UNIR}}, La Rioja, Spain \and
\href{https://www.fidesol.org/}{\texttt{FIDESOL}}, Granada, Spain\\
\email{mcaruso@fidesol.org} \and
\texttt{CONICET} - \texttt{UBA}, \texttt{ICC}, Argentina\\
\email{gbellomo@icc.fcen.uba.ar}
}

\maketitle
\begin{comment}
% Spanish Abstract + keywords

\begin{abstract} El modelado de Relaciones Cuantitativas Estructura-Actividad (\texttt{QSAR}) es fundamental en el descubrimiento de fármacos, pero los métodos clásicos presentan limitaciones al manejar datos de alta dimensionalidad y al capturar interacciones moleculares complejas. Esta investigación propone mejorar las técnicas \texttt{QSAR} mediante Máquinas de Vectores de Soporte Cuánticas (\texttt{QSVMs}), que aprovechan los principios de la computación cuántica para procesar información en espacios de Hilbert. Utilizando codificación cuántica de datos y funciones de núcleo cuántico, buscamos desarrollar modelos predictivos más precisos y eficientes.

\keywords{\texttt{QSAR}, clasificación, descubrimiento de fármacos, máquinas de soporte vectorial, núcleo cuántico.}

\end{abstract}
\end{comment}
\begin{abstract}
Quantitative Structure-Activity Relationship (\texttt{QSAR}) modeling is key in drug discovery, but classical methods face limitations when handling high-dimensional data and capturing complex molecular interactions. This research proposes enhancing \texttt{QSAR} techniques through Quantum Support Vector Machines (\texttt{QSVMs}), which leverage quantum computing principles to process information in Hilbert spaces. By using quantum data encoding and quantum kernel functions, we aim to develop more accurate and efficient predictive models.

\vspace{1em}

\textbf{Keywords:} \texttt{QSAR}, classification, drug discovery, support vector machines, quantum kernel.
\end{abstract}

\begin{comment}

% English Title + Abstract + Keywords (manually formatted)
{\centering
{\Large \textbf{Quantum \texttt{QSAR} for drug discovery}}

\begin{minipage}{0.9\linewidth}
Quantitative Structure-Activity Relationship (\texttt{QSAR}) modeling is key in drug discovery, but classical methods face limitations when handling high-dimensional data and capturing complex molecular interactions. This research proposes enhancing \texttt{QSAR} techniques through Quantum Support Vector Machines (\texttt{QSVMs}), which leverage quantum computing principles to process information in Hilbert spaces. By using quantum data encoding and quantum kernel functions, we aim to develop more accurate and efficient predictive models.

\vspace{0.5em}

\textbf{Keywords:} \texttt{QSAR}, classification, drug discovery, support vector machines, quantum kernel.
\end{minipage}
\par}

\end{comment}

\section{Introduction}

\texttt{QSAR} models aim to establish relationships between the physicochemical properties of compounds and their molecular structures. \cite{hansch1964}
These mathematical models serve as valuable tools in pharmacological studies by providing an \textit{in silico} methodology to test and classify new compounds with desired properties, diminish the need for laboratory experimentation \cite{NatarajanEtAl2025Molecules}. \texttt{QSAR} models are used, for example, to predict pharmacokinetic processes such as absorption, distribution, metabolism, and excretion, \texttt{ADME}, which refers to the processes that describe how a drug or chemical substance moves through and is processed by the body.

In other fields, (\texttt{QSAR}) plays an important role; for example, \textit{in silico} toxicity studies have become fundamental in drug development. A prevalent way in which \texttt{QSAR} is used is in this context of prediction, which helps us understand how we can link toxicity outcomes to the structural properties of specific compounds.

Many models show decent performance throughout their implementations, as they rely on a pipeline that is optimizable and improvable, whereas machine learning methods will always involve a tradeoff between accuracy and interpretability.

\subsection{Evolution of \texttt{QSAR} Modeling Approaches}

Traditionally, \texttt{QSAR} relied on linear regression models, but these were quickly replaced by more sophisticated approaches. Bayesian neural networks emerged as a powerful alternative, demonstrating the ability to distinguish between drug-like and non-drug-like molecules with high accuracy \cite{Ajay1998}. These models showed excellent generalization capabilities, correctly classifying more than 90\% of the compounds in the database %(\texttt{CMC})
while maintaining low false positive rates.

Random forest algorithms have also proven to be effective tools for \texttt{QSAR} modeling \cite{Svetnik2003}. This ensemble method, which combines multiple decision trees, has shown superior performance in predicting biological activity based on molecular structure descriptors. Its advantages include built-in performance evaluation, descriptor importance measures, and compound similarity computations weighted by the relative importance of descriptors.

In general, the process involves three main stages: obtaining a training dataset with measured properties of known compounds, encoding information about the compounds' structure, and building a model to predict properties from the encoded structural data, followed by training the model. \textbf{(1.)} Preprocessing and extraction of molecular descriptors. \textbf{(2.)} Encoding of classical data into quantum states using a feature map. \textbf{(3.)} Classification using support vector machines (\texttt{SVM}) with classical and quantum kernels.

\subsection{General Pipeline}

\textbf{1. Compound Collection and Curation:}  
The process begins with the collection of candidate compounds, either from experimental or theoretical sources. These compounds are curated to ensure suitability for the selected biological target. This step may involve filtering based on physicochemical properties or prior biological knowledge.

\textbf{2. Data Preprocessing and Descriptor Calculation:}  
Regardless of the target, all data undergoes preprocessing to normalize and standardize values. Molecular descriptors (features) are computed for each compound. These may include physicochemical properties (e.g., molecular weight, hydrogen bond donors
/acceptors, rotatable bonds) and structural fingerprints (e.g., MACCs, ECFP).  
Given the constraints of current quantum hardware, dimensionality reduction is often necessary. Techniques such as Principal Component Analysis (PCA) are applied to retain the most informative components while reducing the number of features, thus minimizing the required number of qubits for quantum encoding.

\textbf{3. Dataset Balancing and Partitioning:}  
In this study, the dataset is inherently imbalanced and relatively small. Although advanced balancing techniques (e.g., SMOTE, RandomUndersampling) were not applied, the dataset serves as a practical testbed for rapid experimentation and for evaluating the methodology across different data volumes. For future work, balancing strategies could be incorporated to assess their impact on model performance.

\textbf{4. Classical-to-Quantum Data Mapping:}  
Once the dataset is enriched and preprocessed, classical features are mapped to quantum states using a feature map (e.g., ZZFeatureMap). The number of qubits required is determined by the dimensionality of the reduced feature set. This mapping is a critical step, as it enables the exploitation of quantum space. \cite{Schuld2018Quantum}

\textbf{5. Model Training and Evaluation:}  
The enriched dataset is used to train both classical and quantum models. For quantum models, the support vector machine (SVM) leverages quantum kernels \cite{Li2019QuantumInspiredSVM}, with training and inference performed either on quantum simulators or real quantum processing units (QPUs). The choice of platform and the number of qubits used are dictated by the final feature dimensionality and hardware availability.  

Experiments are typically partitioned into training and test sets, with performance metrics (e.g., accuracy) computed to compare classical and quantum approaches.

\textbf{6. Scalability and Implementation Notes:}  
Current quantum hardware imposes strict limits on the number of qubits and circuit depth, constraining the size and complexity of datasets that can be processed. Execution time and noise are also significant factors, especially when running on real QPUs. These limitations highlight the importance of dimensionality reduction and motivate ongoing research into error mitigation and hybrid quantum-classical workflows.  

While this work focuses on quantum SVMs, alternative quantum approaches such as Variational Quantum Circuits (VQCs) could be explored in future studies to further assess the potential of quantum machine learning in QSAR applications.

This detailed pipeline description aims to clarify the methodological steps, justify key design choices, and provide a foundation for reproducibility and future scalability assessments.

\begin{figure}
    \centering
    \includegraphics[width=0.8\linewidth]{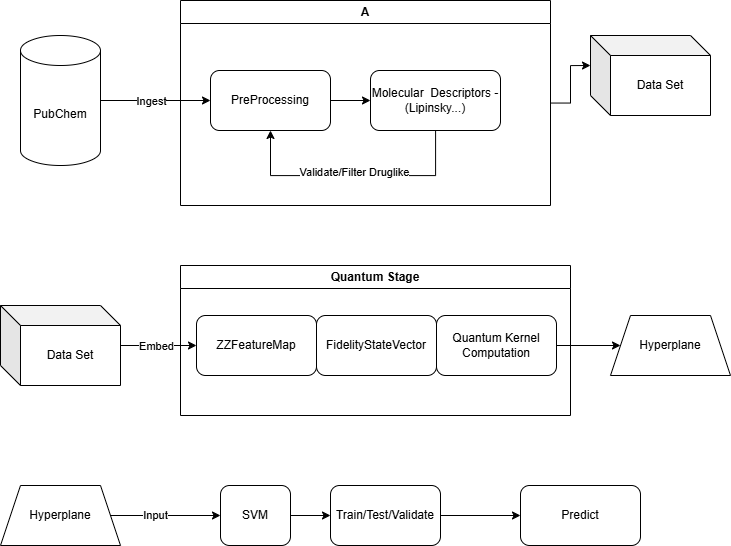}
    \caption{High level pipeline from data perspective}
    \label{fig:enter-label}
\end{figure}

\subsection{Dataset, Descriptors, and Features}

Each candidate molecule has a series of molecular descriptors or \textit{features}, such as the median effective concentration, Lipinski descriptors, %or (\texttt{MACCs}) 
which are a type of molecular fingerprint—specifically a 2D structure fingerprint of $166$ bits—used to represent and compare molecular structures.

These descriptors may include records of experimental results; for example, the concentration $\mathtt{EC}_{50}$ indicates the amount of a compound required to elicit 50\% of the maximum biological effect after a specific exposure time. This is expressed in molar units $M$: $\mathtt{mol}/\mathtt{L}$.

Each dataset requires data processing. In models involving \texttt{ADME}, it is important to work with concentration parameters that provide meaningful information. Concentrations are usually in the $nM$ range, and a logarithmic transformation is employed, defining the potential of this concentration as $p\mathtt{EC}_{50}$, denoted by $p = -\log_{10} \left( \mathtt{EC}_{50} \times 10^{-9} \right)$, which facilitates its use in quantitative analyses of biological activity \cite{Neubig2003}.

To contextualize these models within a study domain, we will use a dataset where the target is the \texttt{M2 Muscarinic Acetylcholine} receptor, a G protein-coupled receptor that plays a crucial role in the parasympathetic nervous system, particularly in regulating cardiac function and smooth muscle activity. It is encoded by the \texttt{CHRM2} gene in humans.

In the pharmacokinetic context, we will use Lipinski's \textit{rule of five}, which is a set of empirical criteria fundamental to drug design. It describes molecular properties relevant to pharmacokinetics in the human body, including absorption, distribution, metabolism, and excretion (\texttt{ADME}). This rule helps assess the likelihood that a chemical compound exhibits adequate pharmacokinetic properties for oral administration in humans, based on four key molecular properties: 
molecular weight ($\leq 500 \mathtt{Da}$), number of hydrogen bond donors ($\leq 5$), number of hydrogen bond acceptors ($\leq 10$), and octanol-water partition coefficient ($\log P\leq 5$). A compound that meets at least three of these criteria is more likely to have good oral bioavailability.

It is important to note that, while this rule is useful for predicting pharmacokinetic properties, it does not predict whether a compound will be pharmacologically active. Its main utility lies in the early stages of drug discovery, allowing researchers to filter out compounds with a low probability of success before conducting costly experiments.

%\subsection{Features, Descriptors}

From the structural information of the molecules, various descriptors of interest are extracted, among which the following are highlighted: number of hydrogen bond donors, $n_d$, representing the number of functional groups that can donate a hydrogen atom; number of hydrogen bond acceptors, $n_a$, which counts the number of sites capable of accepting a hydrogen atom; number of rotatable bonds, $\rho$, indicating molecular flexibility associated with the ability to rotate around single bonds; molecular weight, $w$, which defines the mass of the molecule in atomic mass units.

In this initial study, we consider a limited number of descriptors, which will depend on the experiments described later, associated with the representational power of classical data in quantum systems. For practical purposes, these features were defined as part of a potentially more refined feature engineering process.

\subsection{Data Processing for the Model}

Regarding data processing, we aim to maintain a consistent scale of values to enable operations, standardize values, and in some cases compress data. Therefore, proper preparations will be carried out before training any model. In this way, the feature vector is given by $\pmb{x} = (n_d,n_a,\rho,w,\cdots)$. Due to the variability in numerical scales of these descriptors, normalization is performed using the \texttt{minmax} method, so that each component $l$ of the rescaled vector, $\pmb{x}'$, is expressed as $x'_l=\pmb{(}x_l-\mathtt{min}\{x_l\}\pmb{)}/\pmb{(}\mathtt{max}\{x_l\}-\mathtt{min}\{x_l\}\pmb{)}$, e.g., $\forall l$: $x'_l\in [0,1]$.

\section{Classical and Quantum Models}

In the context of supervised machine learning, we work with labeled data, particularly a training dataset of size $N$, $\{(\pmb{x}_i,y_i)\}_{i\in I_N}$, where $\pmb{x}_i$ is the feature vector and $y_i$ its corresponding label indicating whether it is suitable or not. The goal is to find a predictor for $y$ from a family of parameterized predictors with a real-valued parameter vector $\pmb{q}$, by solving an optimization problem for a function of $\pmb{q}$. Specifically, we consider regression and classification models, with predictor families defined respectively by the following functions $f$ and $g$:
$$f(\pmb{x},\pmb{q})=\sum_{k=1}^{m}q_k \phi_k(\pmb{x}), \quad 
g(\pmb{x},\pmb{\alpha},b)=\mathtt{sgn}\left[\sum_{i=1}^{N}\alpha_i y_iK(\pmb{x}_i,\pmb{x})+b\right]$$ 
where in the second case the parameter vector is $\pmb{q}=(\pmb{\alpha},b)$.
We aim to estimate whether a given compound is suitable using $\hat{y}$, which corresponds to the output of the respective trained predictors. The functions $\phi_k(\pmb{x})$ are called \textit{feature maps}, and they are used to transform the features $\pmb{x}$ to another space—either of lower dimensionality or one that reveals separability between two given points $\pmb{x},\pmb{x}'$. The function $K(\pmb{x},\pmb{x}')$ is called a \textit{kernel}, and its dependence on $(\pmb{x},\pmb{x}')$ arises through the feature maps $(\phi(\pmb{x}),\phi(\pmb{x}'))$ \cite{HuangEtAl2022Science}. Some examples include the linear kernel 
$\phi(\pmb{x})\cdot\phi(\pmb{x}')$, the polynomial kernel $(\phi(\pmb{x})\cdot\phi(\pmb{x}')+c)^d$, or the Gaussian kernel $\exp[-\gamma||\phi(\pmb{x})-\phi(\pmb{x}')||^2]$.

Since the data is classical, quantum computing could add value in two parts of the process: \textbf{1.} solving the optimization problems underlying the training phase, and \textbf{2.} encoding data using quantum kernels. The data is encoded into quantum states through a feature map implemented by a unitary operator $U(\pmb{x})$, giving the representation $\ket{\phi(\pmb{x})} = U(\pmb{x})\ket{0}^{\otimes n}$. The similarity between two quantum states is measured using fidelity, and the quantum kernel is defined as $K_q(\pmb{x},\pmb{x}') = \left| \langle \phi(\pmb{x}) | \phi(\pmb{x}') \rangle \right|^2$. This kernel naturally incorporates superposition and entanglement effects, enabling the capture of complex nonlinear relationships in feature space \cite{HuangEtAl2021NatComm}. 
Classification is then carried out by training an \texttt{SVM} where the classical kernel is replaced by the quantum kernel $K_q$. This approach allows us to explore the efficiency and potential of quantum algorithms in \texttt{QSAR} scenarios, comparing them with classical approaches \cite{HavlicekEtAl2019Nature}.
The underlying optimization problem in both regression and classification models involves a quadratic problem that can be solved using classical algorithms like gradient descent, heuristic methods like simulated annealing and its quantum variant, or by using gate-based quantum computing to implement algorithms such as \texttt{VQE} or \texttt{QAOA}.

\begin{figure}
    \centering
    \includegraphics[width=0.9\linewidth]{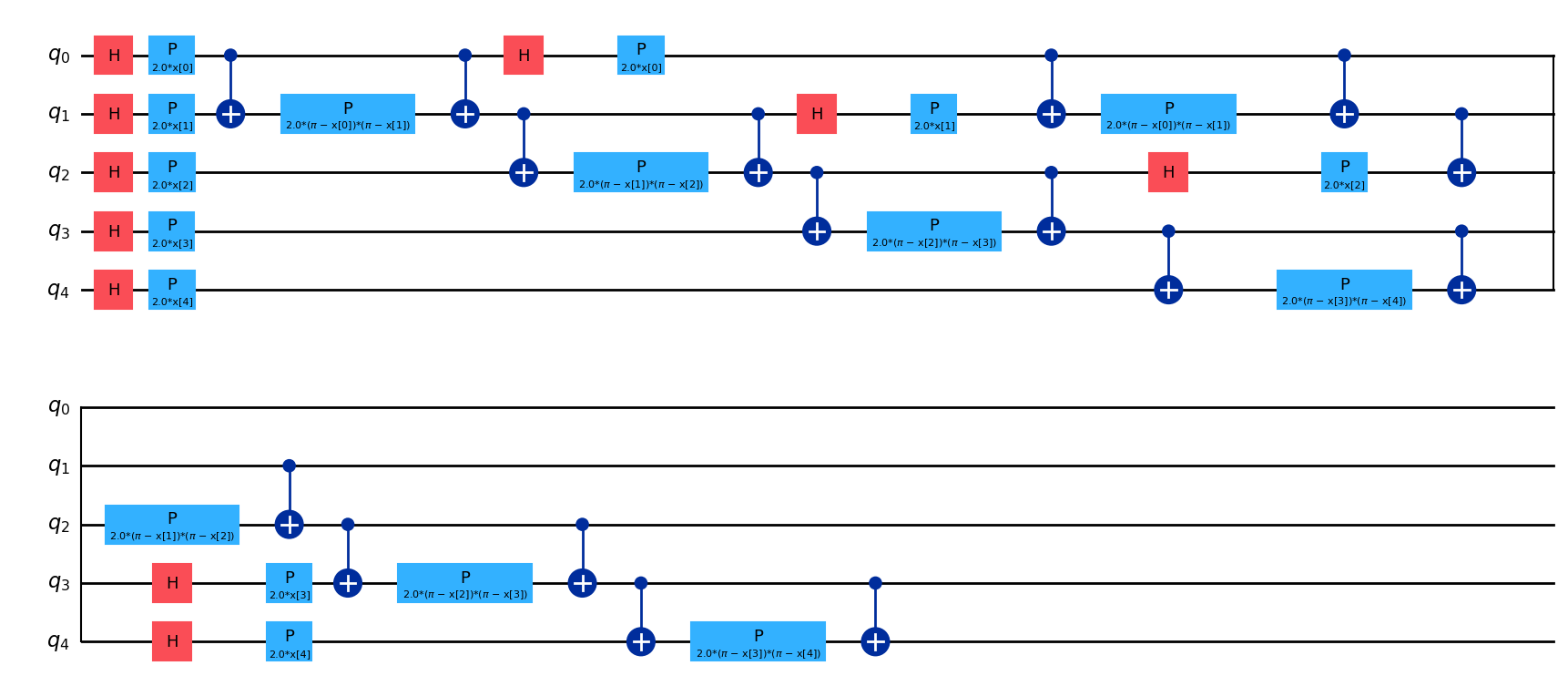}
    \caption{ZZFeaturemap as a linear entangled quantum kernel.}
    \label{fig:enter-label}
\end{figure}

\begin{figure}
    \centering
    \includegraphics[width=0.9\linewidth]{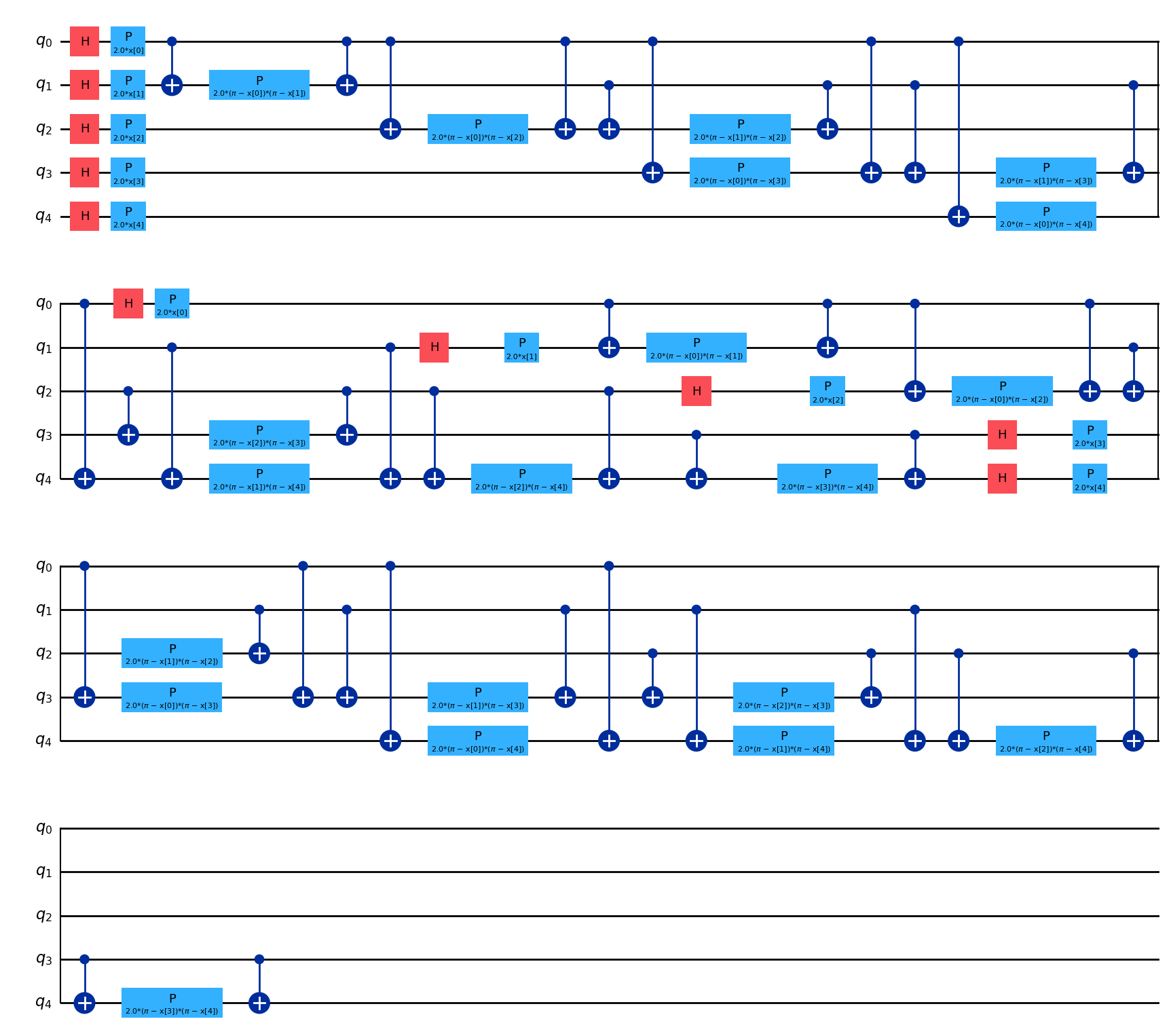}
    \caption{ZZFeaturemap as a full entangled quantum kernel.}
    \label{fig:enter-label}
\end{figure}

\begin{figure}
    \centering
    \includegraphics[width=0.9\linewidth]{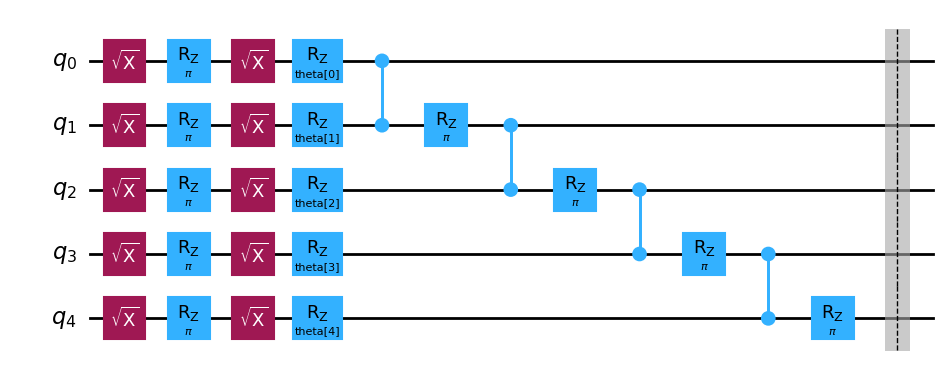}
    \caption{Custom linear entangled quantum kernel.}
    \label{fig:enter-label}
\end{figure}

%%%%%%%%%%%%%

% si ves que no va a haber discusión alguna cambiar el título de la sección a \section{Resultados} 
\section{Results and Discussion}

In this section, we present the results obtained from implementing the regression and classification models. These models can be deployed on either classical or quantum hardware. In particular, for the quantum setting, this includes annealing-based computers such as those from \texttt{DWave}, or gate-based universal quantum computers like those developed by \texttt{IBM}.

To compare the performance across different models, we have chosen the metric known as accuracy, defined as:
\begin{equation}\label{def. accuray}
\mathtt{acc}=\frac{\;1\;}{n} \sum_{i=1}^{n} \pmb{1}[\hat{y}_i=y_i], 
\end{equation}
where $n$ is the number of test samples, and $\pmb{1}[\hat{y}_i=y_i]$ is the indicator function over the set of correct predictions. %, recall that in general, given a set $X$, $\pmb{1}_X:X\rightarrow \{0,1\}$ is such that $\pmb{1}_X(x)=1$ if and only if $x\in X$ and $\pmb{1}_X(x)=0$ otherwise.

A comparison of the different regression (\texttt{REG}) and classification (\texttt{SVM}) models, based on whether classical or quantum algorithms were used, is summarized in Table \ref{tabla}.

\begin{table}[h]
\centering
\caption{We denote by $\pmb{c}$ and $\pmb{q}$ the classical and quantum terms, respectively, to qualify the type of model or the kernel as appropriate. The acronyms \texttt{sim} and \texttt{QPU} refer to execution on quantum simulators and real quantum processors, respectively.}

% REG1 scikitlearn linear reg.
% REG1 QUBO linear reg.
% SVM1 scikitlearn 

\begin{tabular}{c!{\vrule width 1pt}c!{\vrule width 1pt}c!{\vrule width 1pt}c!{\vrule width 1pt}c}
\bottomrule[1pt]
\rowcolor{qnowcolor}
\;\textcolor{white}{\textbf{model}}\;& 
\;\;\textcolor{white}{\textbf{type}}\;\; &
\;\textcolor{white}{$\;\;\mathtt{acc}\;\;$} & 
\;\textcolor{white}{\textbf{execution}} \;& 
\textcolor{white}{\textbf{kernel}} \\
$\mathtt{REG}_1$&$c$& 0.95 & \texttt{CPU} & $-$ \\
\rowcolor{qnowcolor!10}
$\mathtt{REG}_2$& $q$& 0.97 & \texttt{sim} & $-$ \\
$\mathtt{SVM}_1$& $c$ & 0.87 & \texttt{CPU} & $c\;|$ linear\\
\rowcolor{qnowcolor!10}
$\mathtt{SVM}_2$& $c/q$ & 0.98 & \texttt{sim} & $q\;|$ linear - Fig 2 \\
$\mathtt{SVM}_3$& $c/q$ & 0.83 & \texttt{sim} & $q\,|$ nolinear - Fig 3\\
\rowcolor{qnowcolor!10}
$\mathtt{SVM}_4$& $c/q$ & 0.40 & \texttt{QPU} & $q\;|$linear Fig 4\\
\toprule[1pt]
\end{tabular}
\label{tabla}
\end{table}

\section{Conclusions}

A pipeline has been developed that integrates traditional \texttt{QSAR} methods with quantum machine learning techniques. The methodology includes preprocessing and normalization of molecular descriptors, projection of these data into quantum states via the $ZZ$-feature map, and classification using \texttt{SVM} with both classical and quantum kernels. This approach allows for the evaluation of the potential of quantum methods to improve classification in chemico-pharmaceutical applications, relying on rigorous mathematical foundations and the emerging capabilities of quantum computing.

The potential advantages of this integration lie in the ability of quantum kernels to capture complex correlations, even in scenarios with limited data, which may translate into improvements in performance compared to classical techniques.

\section*{Acknowledgment} % Use "Acknowledgments" if using Comsoc mode

This work was supported by the project $\mathtt{ECO-20241014}$ $\textcolor{qnowcolor}{\mathtt{QUORUM}}$ funded by Ministerio de Ciencia, Innovación y Universidades, through $\mathtt{CDTI}$.

% si no usao biblatex debo eliminar esta linea
%\printbibliography 

\bibliography{references}

\begin{thebibliography}{}

\bibitem [\protect \citeauthoryear {%
Ajay%
, Walters%
\BCBL {}\ \BBA {} Murcko%
}{%
Ajay%
\ \protect \BOthers {.}}{%
{\protect \APACyear {1998}}%
}]{%
Ajay1998}
\APACinsertmetastar {%
Ajay1998}%
\begin{APACrefauthors}%
Ajay%
, Walters, W\BPBI P.%
\BCBL {}\ \BBA {} Murcko, M\BPBI A.%
\end{APACrefauthors}%
\unskip\
\newblock
\APACrefYearMonthDay{1998}{}{}.
\newblock
{\BBOQ}\APACrefatitle {Can we learn to distinguish between drug-like and nondrug-like molecules?} {Can we learn to distinguish between drug-like and nondrug-like molecules?}{\BBCQ}
\newblock
\APACjournalVolNumPages{Journal of Medicinal Chemistry}{41}{18}{3314--3324}.
\newblock
\begin{APACrefDOI} \doi{10.1021/jm970666c} \end{APACrefDOI}
\PrintBackRefs{\CurrentBib}

\bibitem [\protect \citeauthoryear {%
Hansch%
\ \BBA {} Fujita%
}{%
Hansch%
\ \BBA {} Fujita%
}{%
{\protect \APACyear {1964}}%
}]{%
hansch1964}
\APACinsertmetastar {%
hansch1964}%
\begin{APACrefauthors}%
Hansch, C.%
\BCBT {}\ \BBA {} Fujita, T.%
\end{APACrefauthors}%
\unskip\
\newblock
\APACrefYearMonthDay{1964}{}{}.
\newblock
{\BBOQ}\APACrefatitle {${\rho}$-${\sigma}$-${\pi}$ Analysis. A Method for the Correlation of Biological Activity and Chemical Structure} {${\rho}$-${\sigma}$-${\pi}$ analysis. a method for the correlation of biological activity and chemical structure}.{\BBCQ}
\newblock
\APACjournalVolNumPages{Journal of the American Chemical Society}{86}{8}{1616--1626}.
\newblock
\begin{APACrefURL} \url{https://doi.org/10.1021/ja01062a035} \end{APACrefURL}
\newblock
\begin{APACrefDOI} \doi{10.1021/ja01062a035} \end{APACrefDOI}
\PrintBackRefs{\CurrentBib}

\bibitem [\protect \citeauthoryear {%
Havl{\'i}{\v c}ek%
\ \protect \BOthers {.}}{%
Havl{\'i}{\v c}ek%
\ \protect \BOthers {.}}{%
{\protect \APACyear {2019}}%
}]{%
HavlicekEtAl2019Nature}
\APACinsertmetastar {%
HavlicekEtAl2019Nature}%
\begin{APACrefauthors}%
Havl{\'i}{\v c}ek, V.%
, C{\'o}rcoles, A\BPBI D.%
, Temme, K.%
, Harrow, A\BPBI W.%
, Kandala, A.%
, Chow, J\BPBI M.%
\BCBL {}\ \BBA {} Gambetta, J\BPBI M.%
\end{APACrefauthors}%
\unskip\
\newblock
\APACrefYearMonthDay{2019}{}{}.
\newblock
{\BBOQ}\APACrefatitle {Supervised learning with quantum-enhanced feature spaces} {Supervised learning with quantum-enhanced feature spaces}.{\BBCQ}
\newblock
\APACjournalVolNumPages{Nature}{567}{7747}{209--212}.
\newblock
\begin{APACrefURL} \url{https://doi.org/10.1038/s41586-019-0980-2} \end{APACrefURL}
\newblock
\begin{APACrefDOI} \doi{10.1038/s41586-019-0980-2} \end{APACrefDOI}
\PrintBackRefs{\CurrentBib}

\bibitem [\protect \citeauthoryear {%
Huang%
\ \protect \BOthers {.}}{%
Huang%
\ \protect \BOthers {.}}{%
{\protect \APACyear {2022}}%
}]{%
HuangEtAl2022Science}
\APACinsertmetastar {%
HuangEtAl2022Science}%
\begin{APACrefauthors}%
Huang, H\BHBI Y.%
, Broughton, M.%
, Cotler, J.%
, Chen, S.%
, Li, J.%
, Mohseni, M.%
\BDBL {}McClean, J\BPBI R.%
\end{APACrefauthors}%
\unskip\
\newblock
\APACrefYearMonthDay{2022}{}{}.
\newblock
{\BBOQ}\APACrefatitle {Quantum advantage in learning from experiments} {Quantum advantage in learning from experiments}.{\BBCQ}
\newblock
\APACjournalVolNumPages{Science}{376}{6598}{1182--1186}.
\newblock
\begin{APACrefURL} \url{https://www.science.org/doi/10.1126/science.abn7293} \end{APACrefURL}
\newblock
\begin{APACrefDOI} \doi{10.1126/science.abn7293} \end{APACrefDOI}
\PrintBackRefs{\CurrentBib}

\bibitem [\protect \citeauthoryear {%
Huang%
\ \protect \BOthers {.}}{%
Huang%
\ \protect \BOthers {.}}{%
{\protect \APACyear {2021}}%
}]{%
HuangEtAl2021NatComm}
\APACinsertmetastar {%
HuangEtAl2021NatComm}%
\begin{APACrefauthors}%
Huang, H\BHBI Y.%
, Broughton, M.%
, Mohseni, M.%
, Babbush, R.%
, Boixo, S.%
, Neven, H.%
\BCBL {}\ \BBA {} McClean, J\BPBI R.%
\end{APACrefauthors}%
\unskip\
\newblock
\APACrefYearMonthDay{2021}{}{}.
\newblock
{\BBOQ}\APACrefatitle {Power of data in quantum machine learning} {Power of data in quantum machine learning}.{\BBCQ}
\newblock
\APACjournalVolNumPages{Nature Communications}{12}{1}{2508}.
\newblock
\begin{APACrefURL} \url{https://doi.org/10.1038/s41467-021-22539-9} \end{APACrefURL}
\newblock
\begin{APACrefDOI} \doi{10.1038/s41467-021-22539-9} \end{APACrefDOI}
\PrintBackRefs{\CurrentBib}

\bibitem [\protect \citeauthoryear {%
Li%
\ \protect \BOthers {.}}{%
Li%
\ \protect \BOthers {.}}{%
{\protect \APACyear {2019}}%
}]{%
Li2019QuantumInspiredSVM}
\APACinsertmetastar {%
Li2019QuantumInspiredSVM}%
\begin{APACrefauthors}%
Li, K.%
\BCBT {}\ \BOthersPeriod {.}
\end{APACrefauthors}%
\unskip\
\newblock
\APACrefYearMonthDay{2019}{}{}.
\newblock
{\BBOQ}\APACrefatitle {Quantum‑Inspired Support Vector Machine} {Quantum‑inspired support vector machine}.{\BBCQ}
\newblock
\APACjournalVolNumPages{arXiv preprint arXiv:1906.08902}{}{}{}.
\newblock
\APACrefnote{arXiv:1906.08902 [cs.LG]}
\PrintBackRefs{\CurrentBib}

\bibitem [\protect \citeauthoryear {%
Natarajan%
, Natarajan%
\BCBL {}\ \BBA {} Basak%
}{%
Natarajan%
\ \protect \BOthers {.}}{%
{\protect \APACyear {2025}}%
}]{%
NatarajanEtAl2025Molecules}
\APACinsertmetastar {%
NatarajanEtAl2025Molecules}%
\begin{APACrefauthors}%
Natarajan, R.%
, Natarajan, G\BPBI S.%
\BCBL {}\ \BBA {} Basak, S\BPBI C.%
\end{APACrefauthors}%
\unskip\
\newblock
\APACrefYearMonthDay{2025}{}{}.
\newblock
{\BBOQ}\APACrefatitle {Quantitative Structure–Activity Relationship ({QSAR}) Modeling of Chiral {CCR2} Antagonists with a Multidimensional Space of Novel Chirality Descriptors} {Quantitative structure–activity relationship ({QSAR}) modeling of chiral {CCR2} antagonists with a multidimensional space of novel chirality descriptors}.{\BBCQ}
\newblock
\APACjournalVolNumPages{Molecules}{30}{2}{307}.
\newblock
\begin{APACrefURL} \url{https://doi.org/10.3390/molecules30020307} \end{APACrefURL}
\newblock
\begin{APACrefDOI} \doi{10.3390/molecules30020307} \end{APACrefDOI}
\PrintBackRefs{\CurrentBib}

\bibitem [\protect \citeauthoryear {%
Neubig%
}{%
Neubig%
}{%
{\protect \APACyear {2003}}%
}]{%
Neubig2003}
\APACinsertmetastar {%
Neubig2003}%
\begin{APACrefauthors}%
Neubig, R\BPBI R.%
\end{APACrefauthors}%
\unskip\
\newblock
\APACrefYearMonthDay{2003}{}{}.
\newblock
{\BBOQ}\APACrefatitle {International Union of Pharmacology Committee on Receptor Nomenclature and Drug Classification. XXXVIII. Update on Terms and Symbols in Quantitative Pharmacology} {International union of pharmacology committee on receptor nomenclature and drug classification. xxxviii. update on terms and symbols in quantitative pharmacology}.{\BBCQ}
\newblock
\APACjournalVolNumPages{Pharmacological Reviews}{55}{4}{597--606}.
\newblock
\begin{APACrefURL} \url{https://doi.org/10.1124/pr.55.4.4} \end{APACrefURL}
\newblock
\begin{APACrefDOI} \doi{10.1124/pr.55.4.4} \end{APACrefDOI}
\PrintBackRefs{\CurrentBib}

\bibitem [\protect \citeauthoryear {%
Schuld%
\ \BBA {} Killoran%
}{%
Schuld%
\ \BBA {} Killoran%
}{%
{\protect \APACyear {2018}}%
}]{%
Schuld2018Quantum}
\APACinsertmetastar {%
Schuld2018Quantum}%
\begin{APACrefauthors}%
Schuld, M.%
\BCBT {}\ \BBA {} Killoran, N.%
\end{APACrefauthors}%
\unskip\
\newblock
\APACrefYearMonthDay{2018}{}{}.
\newblock
{\BBOQ}\APACrefatitle {Quantum machine learning in feature Hilbert spaces} {Quantum machine learning in feature hilbert spaces}.{\BBCQ}
\newblock
\APACjournalVolNumPages{arXiv preprint arXiv:1803.07128}{}{}{}.
\newblock
\APACrefnote{arXiv:1803.07128 [quant-ph]}
\PrintBackRefs{\CurrentBib}

\bibitem [\protect \citeauthoryear {%
Svetnik%
\ \protect \BOthers {.}}{%
Svetnik%
\ \protect \BOthers {.}}{%
{\protect \APACyear {2003}}%
}]{%
Svetnik2003}
\APACinsertmetastar {%
Svetnik2003}%
\begin{APACrefauthors}%
Svetnik, V.%
, Liaw, A.%
, Tong, C.%
, Culberson, J\BPBI C.%
, Sheridan, R\BPBI P.%
\BCBL {}\ \BBA {} Feuston, B\BPBI P.%
\end{APACrefauthors}%
\unskip\
\newblock
\APACrefYearMonthDay{2003}{}{}.
\newblock
{\BBOQ}\APACrefatitle {Random forest: A classification and regression tool for compound classification and QSAR modeling} {Random forest: A classification and regression tool for compound classification and qsar modeling}.{\BBCQ}
\newblock
\APACjournalVolNumPages{Journal of Chemical Information and Computer Sciences}{43}{6}{1947--1958}.
\PrintBackRefs{\CurrentBib}

\end{thebibliography}
\end{document}